\documentclass[10pt,twocolumn,tightenlines,prl,floatfix,superscriptaddress]{revtex4}

\usepackage{epsfig}

\usepackage{graphicx}

\newcommand{\be}{\begin{equation}}
\newcommand{\ee}{\end{equation}}

\begin{document}

\title{\bf Semileptonic $B_{q}\rightarrow D^{\ast}_{q}l\nu $ $(q=s, d, u)$
transitions in  QCD   }
\author{ M. Bayar, K. Azizi}

\affiliation{Physics Department, Middle East Technical University,
06531 Ankara, Turkey }

\begin{abstract}
The form factors relevant to
 $B_{q}\rightarrow D^{\ast}_{q}(J^{P}=1^{-})\ell\nu$ $(q=s, d, u)$ decays are calculated in the
 framework of the three point QCD sum rules approach. The heavy quark
 effective theory prediction of the form factors are obtained.
 The total decay width and branching
 ratio for these decays are also evaluated using the $q^2$ dependencies of these form
 factors.
\end{abstract}

\maketitle
Semileptonic pseudoscalar $B_{q}$ decays are crucial
tools to restrict the Standard Model (SM) parameters and search for
new physics beyond the SM. These decays provide possibility to
calculate the elements of the Cabbibo-Kobayashi-Maskawa (CKM)
matrix, leptonic decay constants as well as the origin of the CP
violation.

$B_{q}\rightarrow D^{\ast}_{q}l\nu$ decas occur via $b\rightarrow c$
transition and form factors are central objects in studying of the
these decays. For the calculation of these form factors,
    we use the QCD sum rules method.

\paragraph{Sum rules for the $B_{q}\rightarrow D^{\ast}_{q}\ell\nu$ transition form factors}
The $B_q \rightarrow  D^{\ast}_{q}$ transitions occur via the
$b\rightarrow c$ transition at the quark level. At this level, the
matrix element for this transition is given by:
\begin{equation}\label{lelement}
M_{q}=\frac{G_{F}}{\sqrt{2}} V_{cb}~\overline{\nu}
~\gamma_{\mu}(1-\gamma_{5})l~\overline{c}
~\gamma_{\mu}(1-\gamma_{5}) b.
\end{equation}
To derive the matrix elements for $B_{q}\rightarrow
    D^{\ast}_{q}l\nu$ decays, it is necessary to sandwich Eq. (\ref{lelement})
between initial and final meson states. The amplitude of the
$B_{q}\rightarrow
    D^{\ast}_{q}l\nu$ decays can be written as follows:
\begin{eqnarray}\label{2au} M&=&\frac{G_{F}}{\sqrt{2}}
V_{cb}~\overline{\nu} ~\gamma_{\mu}(1-\gamma_{5})l \nonumber\\
&&<D^{\ast}_{q}(p',\varepsilon)\mid~\overline{c}
~\gamma_{\mu}(1-\gamma_{5}) b\mid B_{q}(p)>.
\end{eqnarray}

Considering Lorentz and parity invariances, the matrix element
$<D^{\ast}_{q}(p',\varepsilon)\mid\overline{c}\gamma_{\mu}(1-\gamma_{5})
b\mid B_{q}(p)>$ appearing in Eq. (\ref{2au}) can be parameterized
in terms of the form factors below:

\begin{equation}\label{3au}
<D^{\ast}_{q}(p',\varepsilon)\mid\overline{c}\gamma_{\mu} b\mid
B_q(p)>=i\frac{f_{V}(q^2)}{(m_{B_{q}}+m_{D^{\ast}_{q}})}\varepsilon_{\mu\nu\alpha\beta}
\varepsilon^{\ast\nu}p^\alpha p'^\beta,
\end{equation}
\begin{eqnarray}\label{4au}
< D^{\ast}_{q}(p',\varepsilon)\mid\overline{c}\gamma_{\mu}
\gamma_{5} b\mid B_{q}(p)> &=&i\left[f_{0}(q^2)(m_{B_{q}}
+m_{D^{\ast}_{q}})\varepsilon_{\mu}^{\ast}
\right. \nonumber \\
-
\frac{f_{+}(q^2)}{(m_{B_{q}}+m_{D^{\ast}_{q}})}(\varepsilon^{\ast}p)P_{\mu}
&-& \left.
\frac{f_-(q^2)}{(m_{B_{q}}+m_{D^{\ast}_{q}})}(\varepsilon^{\ast}p)q_{\mu}\right],\nonumber \\
\end{eqnarray}

where $f_{V}(q^2)$, $f_{0}(q^2)$, $f_{+}(q^2)$ and $f_{-}(q^2)$ are
the transition form factors and $P_{\mu}=(p+p')_{\mu}$,
$q_{\mu}=(p-p')_{\mu}$. In order to calculate these  form factors,
the QCD sum rules method is applied. Initially the following
correlator is considered:

\begin{eqnarray}\label{6au}
\Pi _{\mu\nu}^{V;A}(p^2,p'^2,q^2)=i^2\int
d^{4}xd^4ye^{-ipx}e^{ip'y}\nonumber \\ \times<0\mid T[J _{\nu
D^{\ast}_{q}}(y) J_{\mu}^{V;A}(0) J_{B_{q}}(x)]\mid  0>,
\end{eqnarray}
where $J _{\nu D^{\ast}_{q}}(y)=\overline{q}\gamma_{\nu} c$ and
$J_{B_{q}}(x)=\overline{b}\gamma_{5}q$ are the interpolating
currents of $D^{\ast}_{q}$ and $B_{q} $ mesons, respectively and
 $J_{\mu}^{V}=~\overline{c}\gamma_{\mu}b $ and $J_{\mu}^{A}=~\overline{c}\gamma_{\mu}\gamma_{5}b$
 are vector and axial vector transition currents.

 From the QCD (theoretical) sides, $\Pi _{\mu\nu}(p^2,p'^2,q^2)$ can
also be calculated by the help of OPE and the double dispersion representation for the
coefficients of corresponding Lorentz structures as:

\begin{equation}\label{10au}
\Pi_i^{per}=-\frac{1}{(2\pi)^2}\int ds'\int
ds\frac{\rho_{i}(s,s',q^2)}{(s-p^2)(s'-p'^2)}+\textrm{ subt. terms.}
\end{equation}

The spectral densities $\rho_{i}(s,s',q^2)$ can be calculated from
the usual Feynman integral with the help of Cutkosky rules which
implies that all quarks are real. After calculations for the
corresponding spectral densities the following expressions are
obtained:

\begin{eqnarray}\label{11au}
\rho_{V}(s,s',q^2)&=&4N_{c}I_{0}(s,s',q^2)\nonumber\\&&\left[{(m_{b}-m_{q})A+(m_{c}-m_{q})B}-m_{q}\right],\nonumber\\
\rho_{0}(s,s',q^2)&=&-2N_{c}I_{0}(s,s',q^2)\Bigg[2m_{q}^{3}-2m_{q}^{2}(m_{c}+m_{b})\nonumber\\
&+&m_{q}(q^{2}+s+s'-2m_{b}m_{c})+[q^{2}(m_{b}-m_{q})\nonumber\\
&+&s(3m_{q}-2m_{c}-m_{b})+s'(m_{q}-m_{b})]A\nonumber\\&+&[q^{2}(m_{c}-m_{q})
+s(m_{q}-m_{c})\nonumber\\&+& s'(3m_{q}-2m_{b}-m_{c})]B
+4(m_{b}-m_{s})C\Bigg],\nonumber \\
\rho_{+}(s,s',q^2)&=&2N_{c}I_{0}(s,s',q^2)\Bigg[m_{q}+(3m_{q}-m_{b})A\nonumber\\&+&(m_{q}-m_{c})B
\nonumber\\&+&2(m_{q}+m_{b})D+2(m_{q}-m_{b})E \Bigg]
,\nonumber \\
\rho_{-}(s,s',q^2)&=&2N_{c}I_{0}(s,s',q^2)\Bigg[-m_{q}+(m_{q}+m_{b})A\nonumber\\&-&(m_{q}+m_{c})B
+2(m_{q}-m_{b})D\nonumber\\&+&2(m_{b}-m_{q})E
\Bigg],\nonumber \\
\end{eqnarray}
where
\begin{eqnarray}\label{12}
I_{0}(s,s',q^2)&=&\frac{1}{4\lambda^{1/2}(s,s',q^2)},\nonumber\\
\lambda(a,b,c)&=&a^{2}+b^{2}+c^{2}-2ac-2bc-2ab,
 \end{eqnarray}

\begin{eqnarray}\label{yeni}
\Delta&=&m_{b}^{2}-m_{q}^{2}-s,\nonumber \\
\Delta'&=&m_{c}^{2}-m_{q}^{2}-s',\nonumber \\
u&=&s+s'-q^{2},\nonumber \\
A&=&\frac{1}{\lambda(s,s',q^{2})}(\Delta' u-2\Delta s),\nonumber\\
B&=&\frac{1}{\lambda(s,s',q^{2})}(\Delta u-2\Delta' s),\nonumber\\
C&=&\frac{1}{2\lambda(s,s',q^{2})}(\Delta'^{2}s+\Delta^{2} s'-
\nonumber\\&&\Delta\Delta' u+m_{q}^{2}(-4s s'+u^{2})),\nonumber\\
D&=&\frac{1}{\lambda(s,s',q^{2})^{2}}[-6 \Delta\Delta' s'
u+\Delta'^{2}(2 s s'+u^{2})
\nonumber\\&+&2s'(3\Delta^{2} s'+m_{q}^{2}(-4s s'+u^{2}))],\nonumber\\
E&=&\frac{1}{\lambda(s,s',q^{2})^{2}}[-3 \Delta^{2}
s'u+2\Delta\Delta'(2 s s'+u^{2}) \nonumber\\&-&u(3\Delta'^{2}
s'+m_{q}^{2}(-4s s'+u^{2}))].
 \end{eqnarray}

 The subscripts V, 0 and $\pm$ correspond to the coefficients of the
 structures proportional to $i\varepsilon_{\mu\nu\alpha\beta}p'^{\alpha}p^{\beta}$, $g_{\mu\nu}$ and $\frac{1}{2}(p_{\mu}p_{\nu}
 \pm p'_{\mu}p_{\nu})$, respectively. In Eq. (\ref{11au}) $N_{c}=3$ is the number of colors.

 The contribution of power corrections, i.e., the contributions of
operators with dimensions $d=3$, $4$ and $5$, are given in Ref.
\cite{bizim}.

 By equating the phenomenological expression and the
OPE expression, and applying double Borel transformations with
respect to the variables $p^2$ and $p'^2$ ($p^2\rightarrow
M_{1}^2,~p'^2\rightarrow M_{2}^2$) in order to suppress the
contributions of higher states and continuum, the QCD sum rules for
the form factors $f_{V}$, $f_{0}$, $f_{+}$ and $f_{-}$ are obtained:
\begin{eqnarray}\label{15au}
f_{i}(q^2)=\kappa\frac{(m_{b}+m_{q})
}{f_{B_{q}}m_{B_{q}}^2}\frac{\eta}{f_{D_{q}^{\ast}}m_{D_{q}^{\ast}}}e^{m_{B_{q}}^2/M_{1}^2+m_{D_{q}^{\ast}}^2/M_{2}^2}
\nonumber
\\\times[\frac{1}{(2\pi)^2}\int_{(m_{c}+m_{s})^{2}}^{s_0'} ds' \int_{f(s')}^{s_0} ds\rho_{i}(s,s',q^2)e^{-s/M_{1}^2-s'/M_{2}^2}\nonumber
\\+\hat{B}(f_{i}^{(3)}+f_{i}^{(4)}+f_{i}^{(5)})],\nonumber\\
\end{eqnarray}
 where $i=V,0$ and $\pm$, and $\hat B$ denotes the double Borel
transformation operator and $\eta=m_{B_{q}}+m_{D_{q}^{\ast}}$ for
$i=V,\pm $ and $\eta=\frac{1}{m_{B_{q}}+m_{D_{q}^{\ast}}}$ for $i=0$
are considered. Here $\kappa=+1$ for $i=\pm$ and $\kappa=-1$ for
$i=0$ and $V$. In Eq. (\ref{15au}), in order to subtract the
contributions of the higher states and the continuum, the
quark-hadron duality assumption is used.

Next, we present the infinite heavy quark mass limit of the form
factors for $B_{q}\rightarrow D^{\ast}_{q}l\nu $ transitions. In
HQET, the following procedure are used (see
\cite{ming,neubert1,kazem}). First, we use the following
parametrization:
 \begin{equation}\label{melau}
 y=\nu\nu'=\frac{m_{B_{q}}^2+m_{D_{q}^{\ast}}^2-q^2}{2m_{B_{q}}m_{D_{q}^{\ast}}}
 \end{equation}
 where $\nu$ and $\nu'$ are
  the four-velocities of the initial and final meson states, respectively  and $y=1$
  is so called zero recoil limit. Next, we try to find the y dependent
  expressions of the form factors by taking
  $m_{b}\rightarrow\infty$, $m_{c}=\frac{m_{b}}{\sqrt{z}}$, where
  z is given by  $\sqrt{z}=y+\sqrt{y^2-1}$ and setting the mass of light quarks to zero.
  In this limit the Borel
  parameters take the form $M_{1}^{2}=2 T_{1} m_{b}$ and $M_{2}^{2}=2 T_{2}
  m_{c}$ where $ T_{1}$ and $ T_{2}$ are the new Borel parameters.

  The new continuum thresholds $\nu_{0}$, and
$\nu_{0}'$ take the following forms in this limit
\begin{equation}\label{17au}
 \nu_{0}=\frac{s_{0}-m_{b}^2}{m_{b}},~~~~~~
 \nu'_{0}=\frac{s'_{0}-m_{c}^2}{m_{c}},
 \end{equation}
 and the new integration variables are defined as:
 \begin{equation}\label{18au}
 \nu=\frac{s-m_{b}^2}{m_{b}},~~~~~~ \nu'=\frac{s'-m_{c}^2}{m_{c}}.
 \end{equation}
 The leptonic decay constants are rescaled:
 \begin{equation}\label{21au}
\hat{f}_{B_{q}}=\sqrt{m_{b}}
f_{B_{q}},~~~~~~~\hat{f}_{D_{q}^{*}}=\sqrt{m_{c}} f_{D_{q}^{*}}.
\end{equation}
After the standard calculations, we obtain the y-dependent
expressions of the form factors as follows:

\begin{eqnarray}\label{22au}
f_{V}&=&\frac{(1+\sqrt{z})}{48
\hat{f}_{D_{q}^{*}}\hat{f}_{B_{q}}z^{1/4}}e^{(\frac{\Lambda}{T_{1}}
+\frac{\overline{\Lambda}}{T_{2}})}\Bigg\{ \frac{3}{\pi^{2}(y+1)
\sqrt{y^{2}-1}}\nonumber\\&&\int_{0}^{\nu_{0}}d\nu\int_{0}^{\nu_{0}'}d\nu'(\nu+\nu')
e^{-\frac{\nu}{2T_{1}}-\frac{\nu'}{2T_{2}}}
\nonumber\\&&\theta(2y\nu\nu'-\nu^{2}-\nu'^2)+16<\overline{q}q>\Bigg[1\nonumber\\&-&\frac{m_{0}^{2}}{8}\Bigg(\frac{1}{2T_{1}^{2}}+\frac{1}{2T_{2}^{2}}
+\frac{1}{3T_{1}T_{2}}(1+\frac{1}{\sqrt{z}}+\frac{1}{z})\Bigg)\Bigg]\Bigg\},\nonumber\\
\end{eqnarray}
\begin{eqnarray}\label{222au}
f_{0}&=&\frac{z^{1/4}}{16
\hat{f}_{D_{q}^{*}}\hat{f}_{B_{q}}(1+\sqrt{z})}
e^{(\frac{\Lambda}{T_{1}}+\frac{\overline{\Lambda}}{T_{2}})}\Bigg\{
\frac{3}{\pi^{2}
\sqrt{y^{2}-1}}\nonumber\\&&\int_{0}^{\nu_{0}}d\nu\int_{0}^{\nu_{0}'}d\nu'(\nu+\nu')
e^{-\frac{\nu}{2T_{1}}-\frac{\nu'}{2T_{2}}}\nonumber\\
&&\theta(2y\nu\nu'-\nu^{2}-\nu'^2)+\frac{<\overline{q}q>\sqrt{z}}{3}\nonumber\\&&\Bigg[
\Bigg(\frac{1}{2}+\frac{1}{2z}+\frac{1}{\sqrt{z}}\Bigg)\Bigg(16-m_{0}^{2}(\frac{1}{T_{1}^{2}}
+\frac{1}{T_{1}^{2}})\Bigg)\nonumber\\
&-& \frac{m_{0}^{2}}{T_{1}T_{2}}
\Bigg(1+\frac{1}{3z^{\frac{3}{2}}}+\frac{4}{3\sqrt{z}}
+\frac{1}{z}+\frac{\sqrt{z}}{3}\Bigg)\Bigg]\Bigg\},
\end{eqnarray}
\begin{eqnarray}\label{2222au}
f_{+}&=&\frac{(1+\sqrt{z})}{96
\hat{f}_{D_{q}^{*}}\hat{f}_{B_{q}}z^{1/4}}e^{(\frac{\Lambda}{T_{1}}
+\frac{\overline{\Lambda}}{T_{2}})}\Bigg\{ \frac{9}{\pi^{2}(y+1)
\sqrt{y^{2}-1}}\nonumber\\
&&\int_{0}^{\nu_{0}}d\nu\int_{0}^{\nu_{0}'}d\nu'(\nu+\nu')
e^{-\frac{\nu}{2T_{1}}-\frac{\nu'}{2T_{2}}}
\nonumber\\
&&\theta(2y\nu\nu'-\nu^{2}-\nu'^2)-16<\overline{q}q>
\Bigg[1\nonumber\\
&+&\frac{m_{0}^{2}}{8}\Bigg(\frac{1}{2T_{1}^{2}}+\frac{1}{2T_{2}^{2}}
+\frac{1}{3T_{1}T_{2}}(1+\frac{1}{\sqrt{z}}+\frac{1}{z})\Bigg)\Bigg]\Bigg\},\nonumber\\
\end{eqnarray}
\begin{eqnarray}\label{22222au}
f_{-}&=&-\frac{(1+\sqrt{z})}{96\hat{f}_{D_{q}^{*}}\hat{f}_{B_{q}}z^{1/4}}
e^{(\frac{\Lambda}{T_{1}}+\frac{\overline{\Lambda}}{T_{2}})}\Bigg\{
\frac{9}{\pi^{2}(y+1)
\sqrt{y^{2}-1}}\nonumber\\
&&\int_{0}^{\nu_{0}}d\nu\int_{0}^{\nu_{0}'}d\nu'(\nu+\nu')e^{-\frac{\nu}{2T_{1}}-\frac{\nu'}{2T_{2}}}
\nonumber\\
&&\theta(2y\nu\nu'-\nu^{2}-\nu'^2)+16<\overline{q}q>\Bigg[1-\nonumber\\
&&\frac{m_{0}^{2}}{8}\Bigg(\frac{1}{2T_{1}^{2}}+\frac{1}{2T_{2}^{2}}
+\frac{1}{3T_{1}T_{2}}(1+\frac{1}{\sqrt{z}}+\frac{1}{z})\Bigg)\Bigg]\Bigg\},\nonumber\\
\end{eqnarray}
where $\Lambda=m_{B_{q}}-m_{b}$ and
$\bar{\Lambda}=m_{D_{q}^{*}}-m_{c}$.

\paragraph{Numerical analysis}
This section is devoted to the numerical analysis for the form
factors $f_{V}(q^2)$, $f_{0}(q^2)$, $f_{+}(q^2)$ and $f_{-}(q^2)$.
The threshold parameters $s_{0}$ and $s_{0}' $ are determined from
the two-point QCD sum rules: $s_{0} =(35\pm 2)~ GeV^2$ \cite{12} and
$s_{0}' =(6-8)~ GeV^2 $ \cite{11}. The Borel parameters $M_{1}^2$
and $M_{2}^2 $ are not physical quantities, hence form factors
should not depend on them. Reliable regions for Borel parameters are
$ 10~ GeV^2 < M_{1}^2 <25~ GeV^2 $ and $ 4~ GeV^2 <M_{2}^2 <10
~GeV^2$.

To determine the decay width of $B_{q} \rightarrow
D_{q}^{\ast}l\nu$, the $q^2$ dependence of the form factors $
f_{V}(q^2)$, $f_{0}(q^2)$, $f_{+}(q^2)$ and $f_{-}(q^2)$ in the
whole physical region $ m_{l}^2 \leq q^2 \leq (m_{B_{q}} -
m_{D_{q}^{\ast}})^2$ are needed. The value of the form factors at
$q^2=0$ are given in Table I.
\begin{table}[h]
\centering
\begin{tabular}{|c||c|c|c|} \hline
$f_{i}(0)$& $B_{s}\rightarrow D_{s}^{\ast}\ell \nu $ &
$B_{d}\rightarrow D_{d}^{\ast}\ell \nu$ & $B_{u}\rightarrow
D_{u}^{\ast}\ell \nu $
\\\cline{1-4}\hline\hline
$f_{V}(0)$ & $0.36\pm0.08$ & $0.47\pm0.13$ & $0.46\pm0.13
$\\\cline{1-4} $f_{0}(0)$ & $0.17\pm0.03$ & $0.24\pm0.05$ &
$0.24\pm0.05$\\\cline{1-4} $f_{+}(0)$ & $0.11\pm0.02$ &
$0.14\pm0.025$ & $0.13\pm0.025$\\\cline{1-4}$f_{-}(0)$
&$-0.13\pm0.03$& $-0.16\pm0.04$ & $-0.15\pm0.04$\\\cline{1-4}
\end{tabular}
\vspace{0.8cm} \caption{The value of the form factors at $q^2=0$}.
\label{tab:2}
\end{table}

\begin{figure}
\begin{center}
\includegraphics[width=6cm]{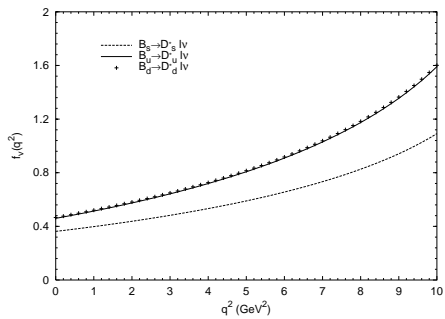}
\end{center}
\caption{The dependence of $f_{V}$ on
 $q^2$ at $M_{1}^2=19~GeV^2$, $M_{2}^2=5~GeV^2$, $s_{0}=35~GeV^2$ and $s_{0}'=6~GeV^2$. } \label{fig1}
\end{figure}

\begin{figure}
\begin{center}
\includegraphics[width=6cm]{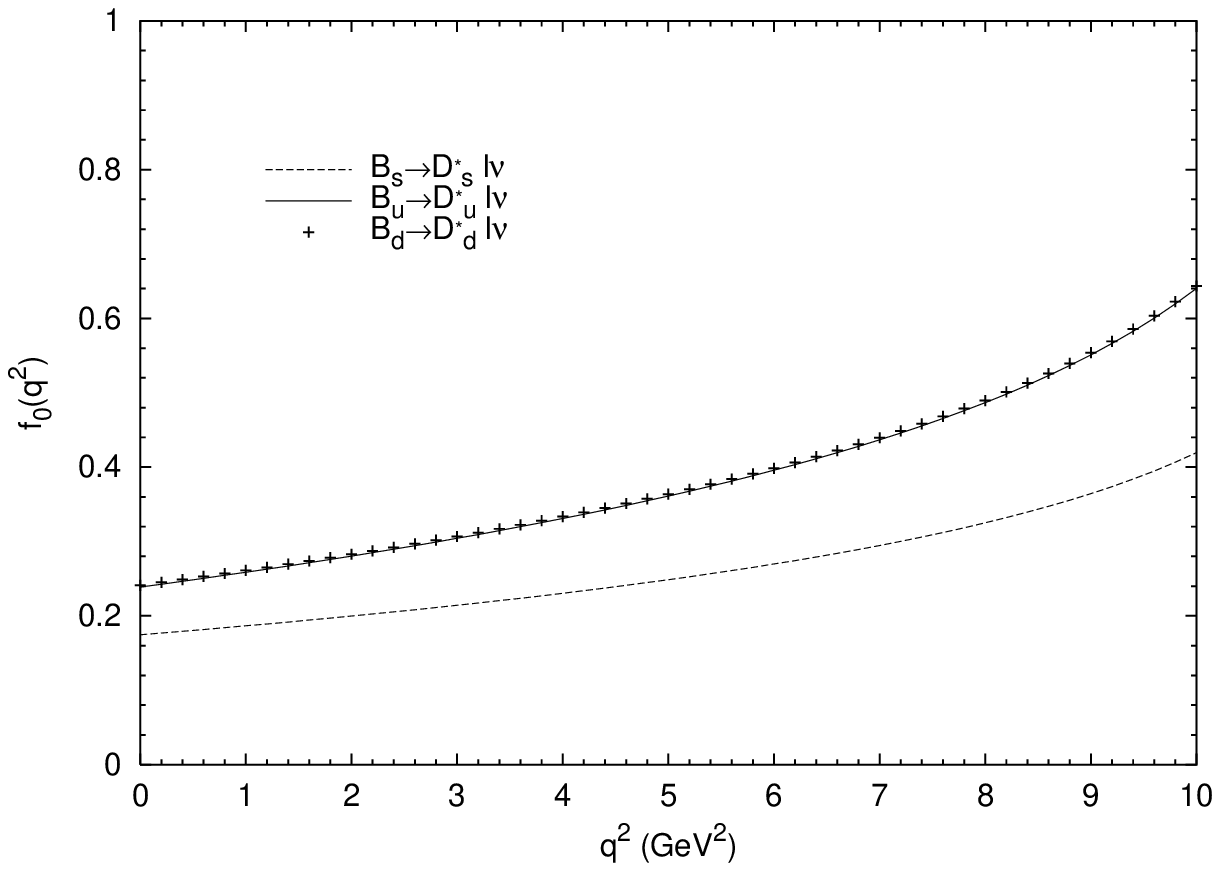}
\end{center}
\caption{ The dependence of $f_{0}$ on
 $q^2$ at $M_{1}^2=19~GeV^2$, $M_{2}^2=5~GeV^2$, $s_{0}=35~GeV^2$ and $s_{0}'=6~GeV^2$.} \label{fig2}
\end{figure}

\begin{figure}
\begin{center}
\includegraphics[width=6cm]{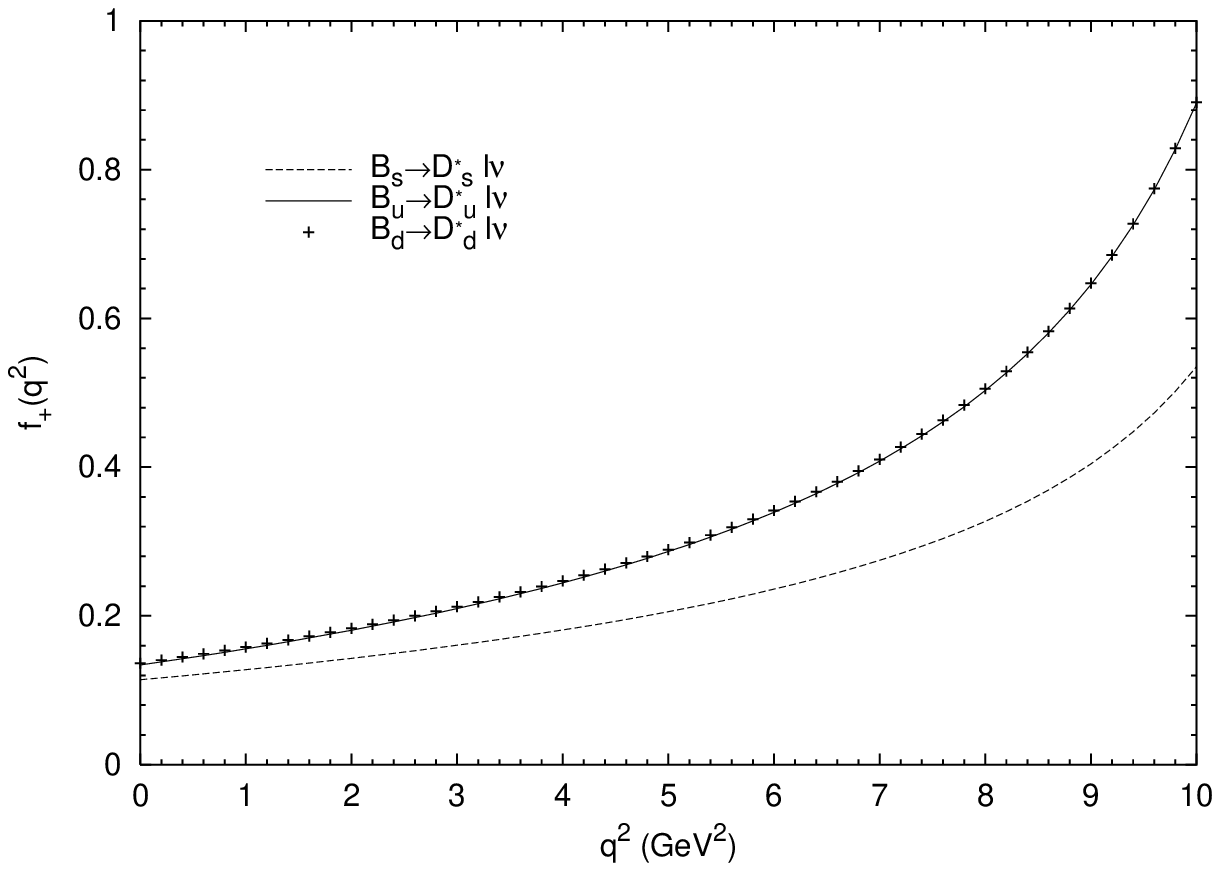}
\end{center}
\caption{The dependence of $f_{+}$ on
 $q^2$ at $M_{1}^2=19~GeV^2$, $M_{2}^2=5~GeV^2$, $s_{0}=35~GeV^2$ and $s_{0}'=6~GeV^2$.} \label{fig3}
\end{figure}

\begin{figure}
\begin{center}
\includegraphics[width=6cm]{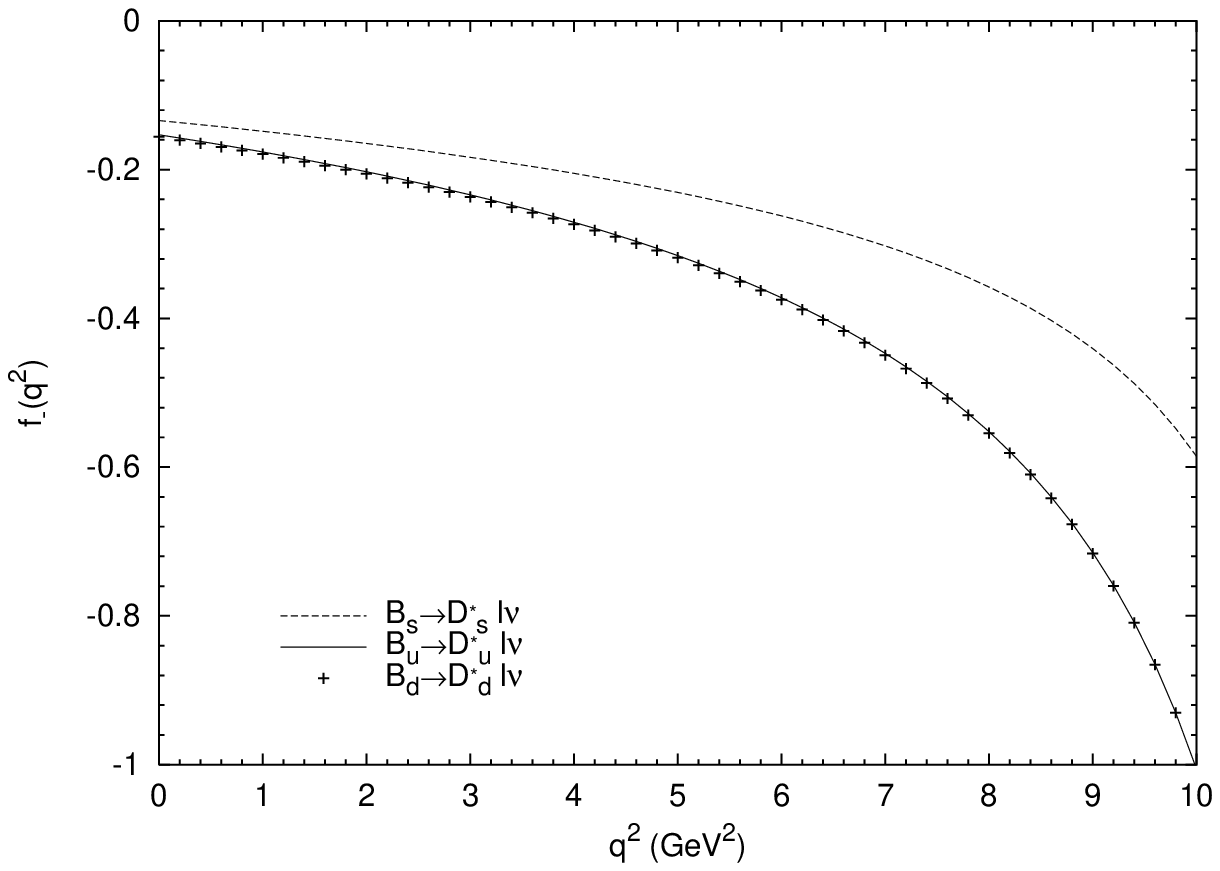}
\end{center}
\caption{The dependence of $f_{-}$ on
 $q^2$ at $M_{1}^2=19~GeV^2$, $M_{2}^2=5~GeV^2$, $s_{0}=35~GeV^2$ and $s_{0}'=6~GeV^2$.} \label{fig4}
\end{figure}

Figs. \ref{fig1}, \ref{fig2}, \ref{fig3} and \ref{fig4} show the
dependence of the form factors $f_{V}(q^2)$, $f_{0}(q^2)$,
$f_{+}(q^2)$ and $f_{-}(q^2)$ on $q^2$. To find the extrapolation of
the form factors, we choose the following fit function.

\begin{equation}\label{17au}
 f_{i}(q^2)=\frac{a}{(q^{2}-m_{B^{*}}^{2})}+\frac{b}{(q^{2}-m_{fit}^{2})}.
\end{equation}
 The values for a, b and
$m_{fit}^{2}$ are given in Table II for example for s case.

\begin{table}[h]
\centering
\begin{tabular}{|c||c|c|c|} \hline
  & a  & b & $m_{fit}^{2}$\\\cline{1-4} \hline \hline
 $f_{V}$ & 55.03 & -54.30 & 23.18\\\cline{1-4}
 $f_{0}$ & 1.43  & -4.32 & 18.80\\\cline{1-4}
 $f_{+}$ & 1.14  & -2.57 & 14.88\\\cline{1-4}
 $f_{-}$ & -2.80  & 3.43 & 14.60\\\cline{1-4}
 \end{tabular}
 \vspace{0.8cm}
\caption{Parameters appearing in the fit function for form factors
of the $B_{s}\rightarrow D_{s}^{\ast}(2112)\ell\nu$ at
$M_{1}^2=19~GeV^2$, $M_{2}^2=5~GeV^2.$} \label{tab:1}
\end{table}

In deriving the numerical values for the ratio of the form factors
at HQET limit, we take the value of the $\Lambda$ and
$\overline{\Lambda}$ obtained from two point sum rules,
$\Lambda=0.62 GeV$ \cite{huang} and $\overline{\Lambda}=0.86
GeV$\cite{dai}. The following relations are defined for the ratio of
the form factors,
\begin{eqnarray}\label{rler}
 R_{1(2)[3]}&=&\Bigg[1-\frac{q^{2}}{(m_{B}+m_{D^{*}})^{2}}\Bigg]\frac{f_{V(+)[-]}(y)}{f_{0}(y)},
\nonumber\\
  R_{4(5)}&=&\Bigg[1-\frac{q^{2}}{(m_{B}+m_{D^{*}})^{2}}\Bigg]\frac{f_{+(-)}(y)}{f_{V}(y)},
  \nonumber\\
  R_{6}&=&\Bigg[1-\frac{q^{2}}{(m_{B}+m_{D^{*}})^{2}}\Bigg]\frac{f_{-}(y)}{f_{+}(y)},\nonumber\\
 \end{eqnarray}

 The numerical values of the above mentioned ratios and a comparison
of our results with the predictions of \cite{neubert2} which
presents the application of the subleading Isgur-Wise form factors
for $B\rightarrow
 D^{\ast}\ell\nu$ are shown in Table III. Note that the values in this
 Table are obtained with $T_{1}=T_{2}=2~ GeV$ correspond to $M_{1}^{2}=19~
 GeV^{2}$ and $M_{2}^{2}=5~ GeV^{2}$.
\begin{table}[h]
\centering
\begin{tabular}{|c||c|c|c|c|c|c|} \hline
  y& 1 (zero recoil) & $1.1$ & $1.2$& $1.3$& $1.4$&1.5\\\cline{1-7}
 $q^2 (GeV^{2})$ & 10.69 & 8.57 & 6.45& 4.33& 2.20&0.08\\\cline{1-7}
  \hline \hline
 $R_{1}$ & 1.34 & 1.31 & 1.25& 1.19& 1.10&0.95\\\cline{1-7}
 $R_{2}$ & 0.80  & 0.99 & 1.10& 1.22& 1.30&1.41\\\cline{1-7}
 $R_{3}$ & -0.80  & -0.79 & -0.80& -0.81& -0.80&-0.80\\\cline{1-7}
 $R_{4}$ & 0.50  & 0.64 & 0.77& 0.94& 1.20&1.46\\\cline{1-7}
 $R_{5}$ & -0.50  & -0.51 & -0.56& -0.62& -0.71&-0.89\\\cline{1-7}
 $R_{6}$ & -0.80  & -0.67 & -0.64& -0.61& -0.55&-0.53\\\cline{1-7}
 $R_{1}$ \cite{neubert2}& 1.31  & 1.30 & 1.29& 1.28& 1.27&1.26\\\cline{1-7}
 $R_{2}$ \cite{neubert2}& 0.90  & 0.90 & 0.91& 0.92& 0.92&0.93\\\cline{1-7}
 \end{tabular}
 \vspace{0.8cm}
\caption{The values for the $R_{i}$ and comparison of $R_{1, 2}$
values with the predictions of \cite{neubert2}.} \label{tab:4}
\end{table}

 The next step is to calculate the differential decay width in terms of the form factors ( see Ref.
\cite{bizim}). The branching ratios are obtained as:
\begin{eqnarray}\label{31au}
\textbf{\emph{B}}(B_s\rightarrow
D_{s}^{\ast}(2112)\ell\nu)&=&(1.89-6.61) \times 10^{-2},\nonumber\\
\textbf{\emph{B}}(B_d\rightarrow
D_{d}^{\ast}(2010)\ell\nu)&=&(4.36-8.94) \times 10^{-2},\nonumber\\
\textbf{\emph{B}}(B_u\rightarrow
D_{u}^{\ast}(2007)\ell\nu)&=&(4.57-9.12) \times 10^{-2}.
\end{eqnarray}

\begin{acknowledgments}
 The authors would like to thank T. M. Aliev and A. Ozpineci for
  their useful discussions and also TUBITAK, Turkish Scientific and Research
Council, for their financial support provided under the project
103T666.
\end{acknowledgments}


\newpage

\end{document}